%% file: paper.tex
\documentclass[preprint,eqsecnum,aps,superscriptaddress]{revtex4}
\usepackage{graphicx}
\usepackage{graphics}
\usepackage{amsmath}  

\newcommand{\ee}{\end{equation}}
\newcommand{\be}{\begin{equation}}
\newcommand{\bea}{\begin{eqnarray}}
\newcommand{\eea}{\end{eqnarray}}
\newcommand{\bml}{\begin{subequations}} 
\newcommand{\eml}{\end{subequations}}

\newcommand{\nuf}{\mbox{\small \boldmath $\nu$}}

\begin{document}
%\title{NDR - Quenching and Current Alteration by Vibronic Effects in Molecular Conductance}
\title{Vibronic effects in single molecule conductance: First-principles 
description and application to benezenealkanethiolates between gold electrodes}

\author{C.\ Benesch}
\affiliation{Department of Chemistry,
  Technical University of Munich,
  Lichtenbergstr.\ 4, D-85747 Garching, Germany}
\author{M.\ \v{C}\'{\i}\v{z}ek}
\affiliation{Charles University, Faculty of Mathematics and Physics, Institute of Theoretical 
Physics, Prague, Czech Republic}
\author{J.\ Klimes}
\affiliation{Charles University, Faculty of Mathematics and Physics, Institute of Theoretical 
Physics, Prague, Czech Republic}
\author{M.\ Thoss}
\affiliation{Department of Chemistry,
  Technical University of Munich,
  Lichtenbergstr.\ 4, D-85747 Garching, Germany}
\author{W.\ Domcke}
\affiliation{Department of Chemistry,
  Technical University of Munich,
  Lichtenbergstr.\ 4, D-85747 Garching, Germany}

\begin{abstract}
The effect of vibrational motion on resonant charge transport through single molecule junctions
is investigated. The study is based on a combination of first-principles electronic structure
calculations
to characterize the system and inelastic scattering theory to calculate transport properties.
The extension of the methodology to describe hole transport through occupied molecular orbitals
is discussed.
The methodology is applied to molecular junctions 
where a benzene molecule is connected via alkanethiolate
bridges to two gold electrodes. The results demonstrate that, depending on the coupling
between the electronic $\pi$-system of the benzene ring and 
the gold electrodes, vibronic coupling may have a significant influence on the
transport properties of the molecular junction.

%We study hole transport through molecular junctions, in which a benzene-ring is bonded to gold
%clusters via alkane thiolate linkers of different size. Our focus is on the vibronic 
%interaction in such
%systems, and we will consider the influence of different contact geometries on the
%I-V characteristics. We use a multi-channel scattering approach in combination with DFT
%calculations to describe the contact and molecule region, and a DFT-parameterized tight-binding
%model of a semi-infinite gold surface to describe the electrodes. Our results show, that vibronic
%interactions are capable of enhancing the current via a shift of electronic resonances to
%stronger coupled positions, and they can lead to quenching of negative differential conductance
%features. Besides the benzene $\pi$-states, which dominate the transport characteristics, we 
%observe a strong indirect influence of sulfur p-states, which interact with the former orbitals
%via the gold contacts.

\end{abstract}

\maketitle

\input{introduction}

\input{theory}

\input{resultsanddiscussion}

\input{conclusions}

\section{Acknowledgment}
We thank Rainer H\"artle and Ivan Kondov for helpful discussions.
This work has been supported by the Deutsche Forschungsgemeinschaft, the German-Israel Science Foundation,
and the Fonds der chemischen Industrie. Support of MC by the 
Alexander von Humboldt foundation and the Grant GACR 202/07/0833 of the Czech grant agency, as well as  
the generous allocation of computing time by the Leibniz Rechenzentrum, Munich,
is gratefully acknowledged.

%\bibliography{Bib/habil,Bib/library,Bib/note}

\end{document}

%% file: introduction.tex
\section{Introduction}

Recent advances in experimental studies of single molecule conduction 
\cite{Reed97,Park00,Smit02,WeberReichert2002,Xiao04,Qiu04,Liu04,Tao2006,Chen07}
have stimulated great interest in the basic mechanisms which 
govern electron transport through nanoscale molecular junctions 
\cite{Haenggi02,Nitzan03,Cuniberti2005}. 
An interesting aspect that distinguishes
molecular conductors from mesoscopic semiconductor devices  is the 
possible influence of the nuclear degrees of freedom 
of the molecular bridge on electron transport.
Due to the small size of molecules, the charging of the molecular bridge
is often accompanied by significant changes of the nuclear geometry.
The current-induced excitation of the vibrations of the molecule
may result in heating of the molecular bridge and possibly breakage of the junction.
Conformational changes of the geometry of the conducting molecule are 
possible mechanisms for switching behavior and negative differential resistance
\cite{Gaudioso2000}. 
Furthermore, the observation of vibrational structures in conduction measurements allows 
the unambiguous identification of the molecular character of the current. 

Vibrational structures in molecular conductance 
were observed, for example, in electron transport experiments on H$_2$, HD, and D$_2$ between 
platinum electrodes \cite{Thijssen2006} using the mechanically controllable 
break junction technique. Thereby, changes in the conductance were 
assigned to switching between two 
different local geometrical configurations induced by the transverse motion of the molecule.
The observed vibrational structures  could be classified as longitudinal or
transversal \cite{Djukic2005}.
In studies of C$_{60}$ molecules between gold electrodes 
\cite{Park00} the center of mass motion of the 
molecular bridge was observed. Other experiments on this system \cite{Boehler07,Parks07} 
as well as on C$_{70}$ \cite{Liu04}, (C$_{70})_2$ \cite{Pasupathy2005}, 
and copper phtalocyanin \cite{Qiu04} on an aluminum oxide film showed structures 
which were related to the internal vibrational modes of the molecule. 
The  aluminum oxide layer  in the latter experiment acted as an insulating layer, which 
effectively reduces the electronic coupling between the 
molecule and the metal substrate thus facilitating the effect of vibrational motion on the conductance.
In another STM experiment \cite{Ogawa2007}, pronounced progressions of vibrational modes were observed
and the dependence of the vibronic coupling on the spatial position of the STM-tip was
demonstrated. 
%Here the lifetime of the anionic state was estimated to be 100\,fs, while the time
%interval between successive tunneling processes was supposed to be 4 orders of magnitude
%smaller, encouraging a single electron transport picture.
Moreover, vibrational signatures of molecular bridges have also been observed in off-resonant 
inelastic electron tunneling spectroscopy (IETS) \cite{Gaudioso2001,Kushmerick04}.

This tremendous experimental progress has inspired great interest in the 
theoretical modelling and simulation
of vibrationally-coupled electron transport in molecular junctions 
\cite{Galperin2007,Gagliardi07,Sergueev07,Pecchia04,ChenZwolak2004,Frederiksen2007,Troisi03,JiangLuo2006,Galperin06,Emberly00,Schoeller01,MayKuhn2006,Lehmann04,Koch05,Nowack05,Ness01}.
In the low-voltage, off-resonant transport regime combinations of electronic-structure calculations
and non-equilibrium Green's function theory (NEGF), employing the self-consistent 
Born approximation (SCBA), 
have been used to investigate vibrational signatures in IETS 
\cite{Pecchia04,ChenZwolak2004,Frederiksen2007}.
Other approaches used to describe inelastic features in the off-resonant region
include scattering theory combined with an expansion of 
either the  molecular Green's function  to first order in vibronic interaction \cite{Troisi03}
or the systems wavefunction in the nuclear coordinates \cite{JiangLuo2006}.

The majority of the studies of vibronic effects
in the resonant tunneling regime (for higher voltages) have been based on generic 
tight-binding models, using an NEGF approach based on the equation of motion method on the
Keldysh contour \cite{Galperin06}, or kinetic rate equations to calculate the current 
\cite{Emberly00,Schoeller01,MayKuhn2006,Lehmann04,Koch05,Nowack05}. 
These model studies have demonstrated that the vibrational
motion of the molecular bridge may affect the current-voltage characteristics significantly.
%A review on vibrational effects in molecular transport junctions is given in 
%Ref.\,\cite{Galperin2007}.

Many of the studies of vibrationally-coupled electron transport 
reported so far invoke approximations which restrict their applicability
either to small electronic-vibrational coupling, small molecule-lead coupling or 
separability of the vibrational modes. 
To circumvent this limitation, 
we recently proposed an approach \cite{Cizek04,Cizek05,note4},
which is based on inelastic multi-channel scattering theory 
and the projection-operator formalism of resonant 
electron-molecule scattering \cite{Domcke91}. 
Within the single-electron description of electron conduction,
this approach is valid for strong electronic-vibrational and molecule-lead coupling and thus 
allows the study of electron transport in the resonant regime.

In previous work \cite{Cizek04,Cizek05}, we have applied this formalism to study vibronic effects
in molecular conductance based on generic models for molecular junctions.
In a recent short communication \cite{Benesch06}, we have combined  the scattering theory approach
with first-principles electronic-structure calculations
to characterize the molecular junction. 
In this paper, we give a detailed 
description of the methodology including a derivation of the model Hamiltonian
for hole transport, a description of the partitioning method used to determine the parameters 
of the Hamiltonian 
by first-principles electronic structure calculations and an outline of the scattering theory approach. 
The methodology is applied to study vibronic effects on conductance 
in molecular junctions where a benzene ring is connected via alkanethiol
bridges to two gold electrodes (Au-S-(CH$_2$)$_n$-C$_6$H$_4$-(CH$_2$)$_n$-S-Au, cf.\ Fig.\ 
\ref{modelsystem}).
This class of molecular junctions was chosen, because
the change of the length of the alkyl chain, $n$, 
allows a systematic variation of the coupling between the electronic $\pi$-system of the phenyl ring and 
the gold electrodes and thus a variation of
the lifetime of the electron on the molecular bridge.
Our previous studies show that, due to the strong molecule lead coupling, 
benzenethiolate (corresponding to $n=0$) 
between gold electrodes exhibits only
minor vibronic effects \cite{Benesch06}.
In this work, we consider benzene-di(ethanethiolate) 
($n=2$) and benzene-di(butanethiolate) ($n=4$),
which correspond to moderate and weak molecule-lead coupling, respectively.

%The paper is organized as follows:
%In section 2 we will describe the details of the transport theory, which, as compared to our
%previous work \cite{Benesch2006}, will be reformulated in order to describe hole transport rather
%than electron transport. The importance of differentiating between the two is elucidated in
%Ref.\,\cite{Galperin06}.
%As we saw in \cite{Benesch2006}, benzenedithiol clamped between gold leads does, because of too
%strong electronic coupling, not show any vibrational effects in the current-voltage
%characteristic. Therefore, in this paper we will focus on the transport behavior of molecules 
%where the electronic coupling between benzene-ring and leads is decreased by either an ethyl- or 
%a butyl-spacer group.
%Thus, in section 3 current-voltage characteristics and transmission
%probabilities of $p$-benzene-di(ethanethiolate) (BDET) with a pyramidal  and a cuboid-shaped contact
%geometry, and of $p$-benzene-di(butanethiolate) (BDBT) with a pyramidal  electrode geometry
%(Fig.\ \ref{modelsystem}) are shown and discussed. Section 4 contains the conclusions.

%% file: theory.tex
\section{Theory}

\subsection{Charge transport model Hamiltonian}

To study charge transport through single molecule junctions employing 
inelastic scattering theory, we use an 
effective single-particle model Hamiltonian with parameters determined by first-principles 
electronic structure calculations. In this section we briefly outline the derivation of the 
Hamiltonian. %A more detailed discussion is given in \ref{Martin08}.
  
The Hamiltonian of a metal-molecule-metal (MMM) junction including the electronic-nuclear
interaction is given by the generic expression (we use atomic units where $\hbar=e=1$),
\begin{equation}
H=T_n+V_{nn}({\bf R})+H_e({\bf R}) .
\end{equation}
Here $T_n$ denotes the kinetic energy of the nuclei, 
\begin{equation}
T_n=\sum_a \frac{{\bf p}^2_a}{2 M_a},
\end{equation}
where
${\bf p}_a$ is the momentum and $M_a$ is the mass of nucleus $a$.
The Coulomb interaction of the nuclei is given by
\begin{equation}
V_{nn}({\bf R})=\sum_{a\neq b}\frac{Z_a Z_b}{|{\bf R}_a-{\bf R}_b|},
\end{equation}
where $Z_a$
is the atomic number and ${\bf R}_a$ denotes the Cartesian coordinates of the position of nucleus $a$.
The electronic Hamiltonian, $H_e({\bf R})$  includes the kinetic energy and 
the Coulomb interaction of the electrons as well as the electronic-nuclear Coulomb 
interaction and depends parametrically on the nuclear coordinates ${\bf R}$.

We first focus on the electronic degrees of freedom and
consider the equilibrium situation without bias voltage at zero temperature.
In this case, the junction is in its electronic ground
state. Employing an effective one particle description, the ground state is given by a single
Slater determinant, $|\Psi_g\rangle=\prod_{i=1}^{n_{el}} c_i^{\dagger}|0\rangle$, where 
$|0\rangle$ denotes the vacuum state, the operator $c_i^{\dagger}$
creates an electron in the single particle state (molecular orbital) $i$ and the product is taken
over all $n_{el}$ electrons in the neutral junction. 
$|\Psi_g\rangle$ is an approximate solution of the 
electronic Schr\"odinger equation,
\begin{equation}
H_e({\bf R})|\Psi_g\rangle= E_g({\bf R})|\Psi_g\rangle
\end{equation}
with eigenenergy $E_g({\bf R})$, which parametrically depends on the nuclear coordinates.
The energy $E_g({\bf R})+V_{nn}({\bf R})$ represents the adiabatic potential energy surface of the electronic 
ground state, which will serve as a reference state in the following. 

If an external voltage is applied to the MMM-junction, electrons (or holes) are transferred
from an external reservoir to the junction and vice versa, i.e.\ the number of electrons on the 
junction, $n_{el}$, will change. Here we consider charge transfer processes which result in  one 
additional electron (corresponding to the anion) or one additional hole 
(corresponding to the cation) on the junction.
Within Koopmans' theorem the 
energy of a cation is given by
subtracting the energy of the single-particle state (molecular orbital),
from which the electron was removed, from the energy of the
neutral ground state. The energy of an anion, on the other hand, is given by adding the energy 
of the single-particle state that was occupied to the energy of the neutral ground state. 
An electronic Hamiltonian that can describe both situations is given by \cite{Domcke1974}
\begin{equation}
H_e({\bf R})=E_g({\bf R})-\sum_{j\in {\rm occ.}} E_j ({\bf R}) c_j c_j^{\dagger}
+\sum_{i \in {\rm unocc.}} E_i({\bf R}) c_i^{\dagger} c_i ,
\label{he_2}
\end{equation}
where the index $j$ labels all orbitals that are occupied in the neutral reference state, while
the index $i$ labels all orbitals that are unoccupied in the neutral reference state. 
Thereby, effects due to electron-electron correlation and orbital relaxation 
have been neglected \cite{note1}.

If the electronic spectrum of a
molecular junction is such that the levels, which contribute to conduction, are all located 
far below or far above the Fermi energy, 
the third or second term in Eq.\ (\ref{he_2}) can be neglected,
respectively. In the latter case the current is dominated by electron transport, while in 
the former case hole transport prevails.
Thus in the two limiting cases, we obtain the electronic Hamiltonian
\bml
\begin{equation}
H_e({\bf R})=E_g({\bf R})+\sum_i E_i ({\bf R}) c_i^{\dagger} c_i ,
\end{equation}
\begin{equation}
H_e({\bf R})=E_g({\bf R})-\sum_j E_j ({\bf R}) c_j c_j^{\dagger}.
\label{he_3}
\end{equation}
\eml
for electron or hole transport, respectively. 
In all applications considered below, the transport is dominated  by 
hole transport. Therefore, we will consider in the following only this case. 
A more detailed discussion and comparison of both cases is given in 
Ref.\ \onlinecite{note_Cizek08}.

To describe charge transport trough a molecular junction it is expedient 
to partition the overall system
into the molecule and the left and right leads. As will be described in more detail in Sec.\ 
\ref{sec_elstruc}, this results in a splitting of the second term in Eq.\ (\ref{he_3}) 
into three terms, describing the molecule, the 
left electrode, and the right electrode separately,
\begin{eqnarray}
H_e({\bf R})&=&E_g({\bf R})-\sum_{m\in M} E_m ({\bf R}) c_m c_m^{\dagger}
-\sum_{k\in L,R} E_k c_k c_k^{\dagger}\nonumber\\
&&-\sum_{m\in M}\sum_{k\in L,R} \left(V_{mk} c_k c_m^{\dagger} + V_{km} c_m c_k^{\dagger} \right) .
\label{he_4}
\end{eqnarray}
Here the indices $m$ and $k$ denote orbitals that are either localized on the molecule (M), or
on the left (L) or right (R) lead, respectively, $c_m^{\dagger}$ and 
and $c_k^{\dagger}$ are the corresponding creation operators, and $V_{mk}$ describes the coupling between 
molecule and leads.
In Eq.\ (\ref{he_4}), we have assumed that direct interactions between left and right lead 
can be neglected and that the energies of the lead states $E_k$, as well as the 
coupling matrix elements, $V_{mk}$, 
are approximately independent on the nuclear coordinates \cite{note3}. 

In the applications considered below, we  use the harmonic approximation for the nuclei.
In this approximation, the potential energy of the nuclei in the electronic ground state, 
$E_g({\bf R})+V_{nn}({\bf R})$, 
%and the molecular energies $E_j({\bf R})$ 
is expanded up to second  order
in the nuclear coordinates around the equilibrium geometry (${\bf R}_{eq}$) of the 
neutral molecule. 
This results in the following expression for the nuclear Hamiltonian of the neutral reference state
\begin{equation}
T_n+V_{nn}({\bf R})+E_g({\bf R})=H_{n0}=\sum_l \omega_l
\left(a_l^{\dagger}a_l+\frac{1}{2}\right), 
\end{equation}
where $\omega_l$ is the frequency of the $l$th vibrational normal mode
of the molecule with  dimensionless coordinate  $q_l$  
and  $a_l^{\dagger}$, $a_l$ denote the  corresponding creation and annihilation operators.
Employing, furthermore, an expansion of the molecular energies $E_j ({\bf R})$ up to first order
around the equilibrium geometry of the 
neutral molecule, the overall Hamiltonian for hole transport is given by
\begin{eqnarray}
H&=&\sum_l \omega_l \left(a_l^{\dagger}a_l+\frac{1}{2}\right)-\sum_{m\in M} 
E_m c_m c_m^{\dagger}
-\sum_{l,m\in M} \frac{\kappa^{(m)}_l}{\sqrt{2}} \left(a_l + a_l^{\dagger}\right) c_m c_m^{\dagger}
\nonumber\\
&&-\sum_{k\in L,R} E_k c_k c_k^{\dagger}
+\sum_{m\in M}\sum_{k\in L,R} \left(V_{mk} c_m^{\dagger} c_k+ V_{km} c_k^{\dagger} c_m\right).
\label{mp_ham}
\end{eqnarray}
Here, we have introduced the electronic-vibrational (vibronic) coupling parameters 
\begin{equation}
\kappa^{(m)}_l=\left(\frac{\partial E_m}{\partial
q_l}\right)_{{\bf R}_{eq}} .
\end{equation}
The linear form of the vibronic coupling implies that only normal modes
belonging to the totally symmetric representation of the respective symmetry group have a
non-vanishing coupling constant. The harmonic approximation of the nuclear potential employed 
in Eq.\ (\ref{mp_ham}) is  valid for small amplitude motion around the 
equilibrium geometry. For a treatment of large amplitude (e.g.\ torsional) motion
in molecular junctions, see Ref.\ \onlinecite{Cizek05}.

The Hamiltonian derived above can be used within different methods to study transport properties
of molecular junctions, e.g. density matrix theory \cite{note_Benesch08} 
or nonequilibrium Green's function approaches
\cite{note_Haertle08}. In this 
paper we use it in combination with inelastic scattering theory. 
When employing scattering theory it is expedient to represent the Hamiltonian in terms of effective
single-particle states for the electronic degrees of freedom.
In the case of hole transport, the relevant single particle states are  
$|\phi_{m} \rangle = c_m|\Psi_g\rangle$, $|\phi_{k} \rangle = c_k|\Psi_g\rangle$, which
describe an additional hole on the molecule or the leads, respectively.
The effective single-particle representation of the Hamiltonian is obtained by projecting
Eq.\ (\ref{mp_ham}) onto the single-particle space, i.e.\ 
$H\rightarrow \sum_{i,i'}|\phi_i\rangle\langle\phi_i|H|\phi_{i'}\rangle\langle\phi_{i'}|$
This results in
\begin{eqnarray}
\label{sp_ham}
H&=&H_{n0}-H_M-H_{ne}-H_L-H_R-V
\end{eqnarray}
with
\bml
\begin{eqnarray}
H_{n0}&=&\sum_l \omega_l \left( a_l^{\dagger}a_l  +\frac{1}{2}\right),\\
H_M&=&\sum_{m\in M}  | \phi_m \rangle E_m \langle\phi_m|,\\
H_{ne}&=&\sum_{l,m\in M} \frac{\kappa^{(m)}_l}{\sqrt{2}}\left(a_l + a_l^{\dagger}\right)
| \phi_m \rangle \langle\phi_m|,\\
H_L+H_R&=&\sum_{k\in L,R} | \phi_k \rangle E_k  \langle\phi_k|,\\
V&=&\sum_{m\in M}\sum_{k\in L,R} \left( |\phi_{m} \rangle V_{km}\langle\phi_k
|+ |\phi_{k} \rangle V_{mk}\langle\phi_m| \right).
\end{eqnarray}
\eml
The negative signs in Eq.\ (\ref{sp_ham}) are due to the fact that we consider hole transport.
The Hamiltonian for electron transport can be derived in a similar way and is explicitly given in 
Ref.\ \onlinecite{note_Cizek08}. It is noted, that for purely electronic transport calculations (i.e.\
without vibronic coupling, $\kappa^{(m)}_l=0$), the different Hamiltonians for electron and hole 
transport give the same current-voltage characteristic. Including vibronic coupling, 
this is, however, no longer the case \cite{note_Cizek08}.

\subsection{First-principles determination of parameters}
\label{sec_elstruc}

The parameters of the Hamiltonian (\ref{sp_ham}) describing the molecular junction
are determined by first-principles electronic structure calculations. To this end,
the overall junction is partitioned into five parts: The molecular 
part, the part of the left and right electrodes that is explicitly included in the
quantum chemistry calculation (in the following referred to as the contacts), 
and the remaining parts 
of the left and right electrodes. The molecule and the contacts form the extended molecule 
and are treated explicitly by quantum chemistry  calculations. The influence of the 
remaining part of the electrodes is taken into account implicitly employing the 
surface self energy of a gold (111) surface.
%, which was determined using a tight-binding model 
%\cite{Martin08}.

\subsubsection{Electronic structure methods}

The parameters for the explicitly calculated part of the molecular junctions
were determined employing electronic-structure calculations
performed with the TURBOMOLE package (V5-7) \cite{Ahlrichs89}
using density functional theory (DFT) with the B3-LYP hybrid functional and the SV(P)
basis set including ECP-60-MWB on the gold atoms. 

%The model systems exhibited either symmetry group C$_i$ or C$_2$, which
%does not only halve CPU-time, but also simplifies the ensuing transport calculations as it
%establishes a left-right symmetry and thus makes the current an odd function of voltage.

The final structure of the three systems considered (cf.\ Fig.\ \ref{modelsystem}) 
was the result of several optimization cycles:
A full geometry optimization of the isolated neutral molecule was followed by a 
full geometry optimization of the system obtained after replacing the hydrogens on sulfur by 
two gold atoms.  Covalent bonding to two gold atoms is the preferred bond formation 
if no symmetry constraints are applied 
\cite{WeberReichert2002,BaschRatner2003}.
Finally, a second metal layer containing five gold atoms was added on both sides
and the geometry of the system 
was again optimized, thereby keeping the internal coordinates of the second gold layer fixed. 
The final geometry of the metal contacts was obtained by increasing
the second layer by several atoms and/or adding two more metal layers, which are cutouts of the 
(111)-plain of the face-centered-cubic lattice of solid gold.
A single point DFT  calculation was performed on this system. The Fermi energy was approximated by 
the average value of the energies of the HOMO and LUMO.

\subsubsection{Partitioning method and electronic parameters}
\label{sec_partitioning}

The electronic structure calculations discussed above result in 
delocalized molecular orbitals for the extended molecule.
%and the converged Kohn-Sham ($h_{KS}$) matrix and the overlap ($S$) matrix of the
%self-consistent-field procedure ($h_{KS} C = S C \epsilon$) are given with respect to an atomic
%orbital basis.
To determine the single-particle states $|\phi_{j} \rangle$,  $|\phi_{k} \rangle$ 
employed in the Hamiltonian (\ref{sp_ham}), 
which are localized on the leads or the molecule, respectively,
and the corresponding electronic energies and coupling parameters, we employ a partitioning
method. This method is based on the following four steps: (i) separation of the
overall Hilbert space into left contact, molecule, and right contact, 
(ii) partitioning of the Hamiltonian according
to this separation, (iii) introduction of a self energy to describe the influence
of the remaining infinite electrodes   (iv) separate diagonalization of the three blocks of the 
partitioned Hamiltonian. The detailed procedure is discussed in the following.

As was mentioned above, in the present paper we work within an effective single-electron picture.
Thus we identify the effective electronic Hamiltonian with the Kohn-Sham matrix, $F$,
of the self-consistent-field DFT calculation. The Kohn-Sham matrix is represented in a 
set of atomic orbitals $|\chi_n\rangle$. 
To separate the overall Hilbert space into three mutually orthogonal subspaces describing
the molecule and the two contacts, the following projection procedure was
employed: First, the set of atomic orbitals of the overall system, $\{|\chi_n\rangle\}$
is divided into three
groups, which are centered at the left contact, the molecule, and the right contact, respectively.
Since it is advantageous to work with orthogonal orbitals \cite{Kurnikov96,Galperin02},
the set of atomic orbitals of the overall system  is  orthogonalized according to L\"owdin
\cite{Loewdin50,Mayer02}
\begin{eqnarray}
    |\tilde\chi_n \rangle =
    \sum\limits_n (S^{-1/2})_{nl} | \chi_l \rangle~,
\label{eq:lowdin}
\end{eqnarray}
where $S$ denotes the atomic orbital overlap matrix with elements 
$S_{ln}=\langle\chi_l|\chi_n\rangle$.
The new basis functions  obtained, $ |\tilde\chi_n\rangle $, 
exhibit a minimal deviation from the 
original ones in a least-square sense and hence, their  localization is preserved. 
In particular, 
the classification into the three groups (left contact, molecule, right contact) 
is still valid.  
%In contrast to other orthogonalization procedures, such as the
%Gram--Schmidt method, the symmetric orthogonalization according to 
%L\"owdin has also the advantage that it is 'least biased'.

The new set of orthogonal basis functions $\{|\tilde\chi_n\rangle\}$ is then used to partition the Kohn--Sham matrix 
from the converged DFT calculation into the three  subspaces. 
The  Kohn--Sham matrix in the orthogonal basis is given by
\begin{equation}
\tilde F = S^{-\frac{1}{2}} F S^{-\frac{1}{2}}.
\end{equation}
According to the separation of the orthogonalized basis set to the three 
groups -- left contact, molecule, right contact -- the  
transformed Kohn-Sham matrix can be arranged in the following block structure
\begin{equation}
\tilde F =\left( \begin{array}{ccc}
	\tilde F_L 	& \tilde F_{LM} & \tilde F_{LR}\\
	\tilde F_{ML} 	& \tilde F_M	 & \tilde F_{MR}\\
	\tilde F_{RL}  & \tilde F_{RM} & \tilde F_R
\end{array} \right) .
\end{equation}

The thus obtained Kohn-Sham matrix describes the extended molecule, which includes  a  
part of the electrodes. To incorporate the influence
of the remaining (infinite) parts of the electrodes, we use a surface Green's function technique.
Within this method, the influence of the remaining part of the electrodes
 is described by the corresponding 
self energy. Within the Green's-function formalism employed below to calculate transport properties
this amounts to adding the self energy to the Kohn-Sham matrix elements belonging to the gold atoms
of the outermost layer of the contacts 
%At this stage we added a self-energy to the Kohn-Sham matrix elements belonging to the gold atoms
%of the outermost layer of the contacts in order to describe the effect of infinite leads,
\begin{equation}
F^{\rm tot} = \tilde F+\Sigma^{{\rm sf}} .
\end{equation}
The surface self energy $\Sigma^{{\rm sf}}$ was obtained by calculating the surface 
Green's function for a (111) gold surface representing the semi-infinite electrode
\cite{Klimes}. The 
surface  Green's function of a gold (111) surface  was determined using a generalization of 
the method developed by Sancho et al. \cite{Lopez85}. Thereby, the (111) surface of gold was described 
by a tight binding model \cite{Mehl96}.
Thereby, we have  neglected nondiagonal contributions, which correspond to  
coupling between different gold atoms. It is noted that due to the added self energy, 
the Kohn-Sham matrix is no longer real but complex symmetric

In the final step, the three different blocks of the Kohn-Sham matrix  $F^{\rm tot} $
are diagonalized separately.
Denoting the matrices of right eigenvectors for the three subspaces as
$U_L$, $U_M$, $U_R$, respectively, the Kohn-Sham matrix is transformed to the new basis of
eigenvectors
\begin{equation}
 \bar F=U^T  F^{\rm tot} U 
=\left( \begin{array}{ccc}
	E_L		& V_{LM} 	& V_{LR}\\
	V_{ML} 	& E_M		& V_{MR}\\
	V_{RL}  & V_{RM} 	& E_R
\end{array} \right),
\end{equation}
where
\begin{equation}
U=\left( \begin{array}{ccc}
	U_L	& 0 	& 0\\
	0 	& U_M	& 0\\
	0   & 0 	& U_R
\end{array} \right) .
\end{equation}
The elements of the thus transformed Kohn-Sham matrix $\bar F$
determine the electronic parameters of the model Hamiltonian (\ref{sp_ham}):
The diagonal submatrices $E_L$ and $E_R$ contain the lead energies $E_k$, the diagonal submatrix 
$E_M$ contains the eigenenergies of the molecular bridge and the submatrices 
$V$ describe the electronic coupling matrix elements $V_{km}$ 
between states localized on the leads and the molecule.
In all our model systems, the
distance between the two metal clusters is relatively large, Therefore, 
direct lead-lead coupling was small enough to be neglected.

The electronic states in the Hamiltonian (\ref{sp_ham}) can be identified with the eigenstates of 
the three different blocks of the Kohn-Sham matrix and are related to the 
original atomic orbital basis via
\begin{equation}
|\phi_{\alpha}\rangle = \sum_{n,l} U_{\alpha n}(S^{-1/2})_{nl}|\chi_l\rangle,
\end{equation}
where $\alpha \in {L,M,R}$.
While the eigenvectors of $F_M^{\rm tot}$, $| \phi_m \rangle$, 
form an orthogonal basis in molecular space,
the eigenvectors of the non-Hermitian matrix $F_L^{\rm tot}$, $| \phi_k \rangle$,  
form a bi-orthogonal 
basis in the left  lead space. Their overlap, 
$\langle\phi_k|\phi_{k'}\rangle = S_k^U \delta_{kk'}$, where $|\phi_{k}\rangle$ is the right 
eigenvector of $F_L^{\rm tot}$ and $\langle\phi_k|$ is its transposed, is in general a complex number. 
The same is true for the right lead space.
We note that a similar separation scheme was employed in the study of electron transfer dynamics in
dye-semiconductor systems \cite{Kondov2007}.

\subsubsection{Nuclear parameters}

To characterize the nuclear degrees of freedom of the molecular bridge and determine the 
corresponding parameters in the Hamiltonian (\ref{sp_ham}) a normal mode analysis
on an extended molecule including two gold layers (i.e. five gold atoms on each side)  
was performed. Thereby, the atomic mass of the gold atoms 
was set to  $10^9$ atomic mass units to separate the nuclear motion in the molecular
bridge from that of 
the gold clusters.

The electronic-nuclear coupling constants $\kappa^{(m)}_l$ were obtained from the numerical gradients 
of the energies $E_{j}$ with respect to the dimensionless normal coordinates $q_l$,
\begin{equation}
\kappa^{(m)}_l=\frac{E_m(+\Delta q_l)-E_m(-\Delta q_l)}{2 \Delta q_l}.
\label{kappa}
\end{equation}
To this end, two DFT calculations with molecular geometries elongated by $\pm \Delta q_l = \pm  0.1$
from the equilibrium geometry were performed.  The value of $\Delta q_l =  0.1$ was chosen based on convergence tests.
%Test calculations  show that this
%value is large enough to avoid numerical problems, but also small enough so that the difference 
%quotient is still a good approximation to the differential quotient.
%%The results of both DFT calculations entered the projection procedure and the
%vibronic coupling parameters were calculated as follows,
%%

\subsection{Observables of Interest}

Several observables are of interest to characterize the effects of vibrational motion on charge
transport through single molecule junctions.
Here, we will focus on the inelastic transmission probability, which describes the
transmission process of a single
electron or hole through the molecular junction, and the current-voltage characteristics.
In all the examples considered below, the transport is dominated by hole transport.
The inelastic transmission probability of a single hole 
from the left to the right lead
as a function of initial and final energy of the hole is given by the expression
\bea
T_{R\leftarrow L}(E_i,E_f)&=&
4\pi^2 \sum_{ \nuf_i, \nuf_f} 
\sum_{k_i \in L}\sum_{k_f \in R} 
P_{\nuf_i}~\delta(E_{\nuf_f}-E_f-E_{\nuf_i}+E_i)~ \nonumber\\ 
&&\times \delta(E_i-E_{k_i})   \delta(E_f-E_{k_f})
 \left| \langle \nuf_f|\langle k_f|VG(E)V|k_i\rangle |\nuf_i\rangle
\right|^2.
\label{t1}
\eea
Here, the $\delta$-function accounts for energy conservation,
$E\equiv E_{\nuf_i}-E_i = E_{\nuf_f}-E_f$ with $E_{\nuf_i}$, and $E_{\nuf_f}$ being the 
energy of the initial and final vibrational states $|\nuf_i\rangle$,  
$|\nuf_f\rangle$, respectively, $-E_i$ and $-E_f$ are the initial and final energies of the
hole, $G(E) = (E^+ - H)^{-1}$ is the Green's function, and
$P_{\nuf_i} = \langle \nuf_i| \rho_0 |\nuf_i\rangle$ denotes 
the population probability of the initial vibrational 
state, $\rho_0 = e^{-H_{n0}/(k_BT)}/Z$. 
In the systems considered in this work, the hole couples primarily to modes with 
relatively high frequencies. As a result, thermal effects are not expected to be of relevance
and the initial vibrational state is assumed to be the ground state, i.e. 
$\rho_0 = |{\bf 0}\rangle \langle {\bf 0}|$.
The total transmission probability, $T_{R\leftarrow L}(E_i)$, 
is obtained by integrating $T_{R\leftarrow L}(E_i,E_f)$ 
over the total range of final energies of the hole.
\begin{equation}
T_{R\leftarrow L}(E_i)=\int {\rm d} E_f T_{R\leftarrow L}(E_i,E_f)~.
\end{equation}

To calculate the transmission probability, the Green's function $G(E)$
in the expression for the transmission probability
is projected  onto the molecular space using the projection operators
$P = \sum_{m\in M} |\phi_{m}\rangle\langle\phi_{m}|$, 
$Q_L = \sum_{k \in L} |\phi_k\rangle\langle\phi_k|$,
$Q_R = \sum_{k \in R} |\phi_k\rangle\langle\phi_k|$,
as well as the Lippmann-Schwinger equation $G=G_0+G_0 V G$. This results in the expression 
\bea
\label{t2}
T_{R\leftarrow L}(E_i,E_f) &=&
\sum_{ \nuf_i, \nuf_f}P_{\nuf_i}\delta(E_{\nuf_f}-E_f - E_{\nuf_i}+E_i)\\
&& \times {\rm tr_M} \left\{ \langle \nuf_i| \Gamma_L(-E_i)~ G_M^{\dagger}(E_{\nuf_i}-E_i)~ 
|\nuf_f \rangle \langle \nuf_f| ~ \Gamma_R(-E_f)~ G_M(E_{\nuf_i}-E_i) |\nuf_i \rangle \right\} \nonumber,
\eea
where the trace is taken over the electronic states on the molecule (M) and
the Green's function projected onto the molecular bridge is given by
\begin{eqnarray}
G_M(E) &=& PG(E)P \\
&=& \left[ E^+ - PHP 
- \Sigma_L^{{\rm lead}}(E-H_{n0}) - \Sigma_R^{{\rm lead}}(E-H_{n0})\right]^{-1}\nonumber\\
&=& \left[ E^+ + H_M - H_{n0} + H_{ne}
- \Sigma_L^{{\rm lead}}(E-H_{n0}) - \Sigma_R^{{\rm lead}}(E-H_{n0})\right]^{-1} \nonumber
\end{eqnarray} 
Here, $\Sigma_{L/R}^{{\rm lead}}(E)$ denote the self energy due to coupling to the left and right 
lead, respectively
\bml
\begin{eqnarray}
\Sigma_L^{{\rm lead}}(E) &=& PVQ_L(E^+ + H_L)^{-1}Q_LVP = -\frac{i}{2} \Gamma_L(E) +
\Delta_L(E),\\
\Sigma_R^{{\rm lead}}(E) &=& PVQ_R(E^+ + H_R)^{-1}Q_RVP = -\frac{i}{2} \Gamma_R(E) +
\Delta_R(E),
\label{self_energy}
\end{eqnarray}
\eml
where $\Gamma_L(E)$ and $\Delta_L(E)$ are the corresponding width and level-shift functions.
The matrix elements of the self energy in the molecular space are given by
\begin{equation}
(\Sigma_L^{{\rm lead}}(E))_{mn}
=\sum_k \frac{V_{km}  V_{nk} }{(E^++E_k)}.
\end{equation}
%\bea
%\label{t2}
%&&T_{R\leftarrow L}(E_i,E_f) =
%\sum_{\nuf_f}{\rm tr} \left\{\vphantom{G_M^{\dagger}}
%\delta(E_{\nuf_f}-E_f - H_{n0}+E_i)\rho_0 \right.\\
%&& \times \left. \Gamma_L(E_i)~ \tilde{G}_M^{\dagger}(H_{n0}-E_i)~ |\nuf_f\rangle \langle \nuf_f| 
%\Gamma_R(E_f)~ \tilde{G}_M(H_{n0}-E_i) \right\} \nonumber,
%\eea
%%
%with the Green's function, $\tilde{G}_M(H_{n0}-E_i)=-G_M(E_i)$, projected onto the molecular bridge
%%
%\begin{eqnarray}
%G_M(E) &=& PG(E)P = \left[ E^+ - PHP 
%- \Sigma_L^{{\rm lead}}(H_{n0}-E) - \Sigma_R^{{\rm lead}}(H_{n0}-E)\right]^{-1}\nonumber\\
%\tilde{G}_M(H_{n0}-E) &=& \left[ H_{n0}-E-H_{ne}-H_M 
%- \Sigma_L^{{\rm lead}}(H_{n0}-E) - \Sigma_R^{{\rm lead}}(H_{n0}-E)\right]^{-1}\nonumber\\
%\Sigma_L^{{\rm lead}}(E) &=& PVQ_L(E^+ - H_L)^{-1}Q_LVP = -\frac{i}{2} \Gamma_L(E') +
%\Delta_L(E'),\\
%\Sigma_R^{{\rm lead}}(E) &=& PVQ_R(E^+ - H_L)^{-1}Q_RVP = -\frac{i}{2} \Gamma_R(E') +
%\Delta_R(E')
%\label{self_energy}
%\end{eqnarray}
%%
It should be emphasized that the thus defined self energy $\Sigma_L^{{\rm lead}}(E)$ corresponds
to hole transport. It is related to the commonly used self energy for electron transport, 
$\Sigma_{{\rm el}\, L}^{{\rm lead}}(E)$ via
\begin{equation}
(\Sigma_L^{{\rm lead}}(E))_{mn}
= - \left(\Sigma_{{\rm el}\, L}^{{\rm lead}}(-E))_{mn}\right)^\ast .
\end{equation}
In the practical calculations, the latter self energy is obtained  using the partitioning method
outlined above and is explicitely given by
\begin{equation}\label{sigma_matrix}
(\Sigma_{{\rm el}\, L}^{\rm lead}(E))_{mn}=\sum_k \frac{V_{mk}  V_{kn} }{S_k^U (E^+-E_k)} 
\end{equation}
and analogously for $\Sigma_{{\rm el}\,R}^{{\rm lead}}$.
The appearance of the overlap factor $S_k^U$ in Eq.\ (\ref{sigma_matrix}) is a consequence 
of  the bi-orthogonal basis on the leads (cf. the discussion in Sec.\ \ref{sec_partitioning}).
%As a consequence of the bi-orthogonal basis on the leads, an 
%element of this self-energy is given by the expression,
%\begin{equation}
%\Sigma_L^{{\rm lead}}(E)_{ij}=\sum_k \frac{V_{ki}  V_{jk} }{S_k^U (E^++E_k)} ~.
%\end{equation}

It is noted, that in contrast to purely electronic transport calculations  as well as 
applications of the non-equilibrium Green's function formalism to vibronic transport 
\cite{Pecchia04,Ryndyk06}, the Green's function $G_M$ and the self energies 
$\Sigma_L^{\rm lead}$, $\Sigma_R^{\rm lead}$ in Eq.\ (\ref{self_energy}) are operators with 
respect to both the electronic and nuclear degrees of freedom. Thus, the Green's function has to be 
evaluated in the combined electronic-vibrational Hilbert space.
In the calculations reported below, the Green's function $G_M$ is obtained by inverting 
a basis representation of the operator $E^+ - PHP 
- \Sigma_L^{{\rm lead}}(E-H_{n0}) - \Sigma_R^{{\rm lead}}(E-H_{n0})$
for each energy $E$ employing a harmonic  oscillator basis for the vibrational modes.

Based on the transmission probability (Eq.\,\ref{t2}), the current through the molecular junction 
is obtained using a generalized Landauer formula \cite{Nitzan2001}
\begin{eqnarray}\label{g_landauer}
I=\frac{2e}{h}\int {\rm d} E_i \int {\rm d} E_f
\left\{T_{R\leftarrow L} (E_i,E_f) f_R(E_f)[1-f_L(E_i)]\right.\nonumber\\
-\left. T_{L\leftarrow R} (E_i,E_f) f_L(E_f)[1-f_R(E_i)]\right\},
\label{current}
\end{eqnarray}
where $f_L(E)$, $f_R(E)$ denote the Fermi distribution for the left and the right
lead, respectively, taken at zero temperature.

While equations (\ref{t1}) and (\ref{t2}) for the single-hole transmission probability involve no 
approximation, expression (\ref{current}) for the current is valid if many-electron processes 
are negligible. In particular, non-equilibrium effects in the leads and electron correlation due 
to electronic-vibrational coupling are not taken into account. 
Furthermore, it is implicitly assumed, that the nuclear degrees of freedom of the molecular 
bridge relax to the vibrational equilibrium state after scattering of a hole. 
For the applications considered below, comparison
with calculations based on nonequilibrium Green's function methods indicate that these assumptions
are justified \cite{note_Haertle08}.
Without vibronic coupling the expression for the current, Eq.\ (\ref{g_landauer}), reduces
to the usual Landauer formula \cite{Meir92}. 

In principle, the basis states 
$| \phi_{m} \rangle$, $| \phi_k \rangle$, the hole energies 
and the nuclear parameters depend on the bias voltage. 
For the studies in this work, we did not include the voltage self-consistently in the DFT
calculation but used, 
for simplicity, parameters obtained from electronic-structure calculations
at equilibrium and assumed that the bias voltage $V$
enters the formulas only via the chemical potentials of the leads $\mu_{L/R}=\epsilon_f \pm
eV/2$, where $\epsilon_f$ denotes the Fermi energy.
The energies of the lead states for finite voltage are thus given by $E_k \pm eV/2$.
Since we do not invoke the wide-band approximation, the Green's function ${G}_M$, the 
self energies 
$\Sigma_{L/R}^{\rm lead}$ as well as the width functions $\Gamma_{L/R}$ 
also depend on the bias voltage.

%% file: resultsanddiscussion.tex
\section{Results and Discussion}

\begin{figure}
\begin{center}
\includegraphics[ width=8cm]{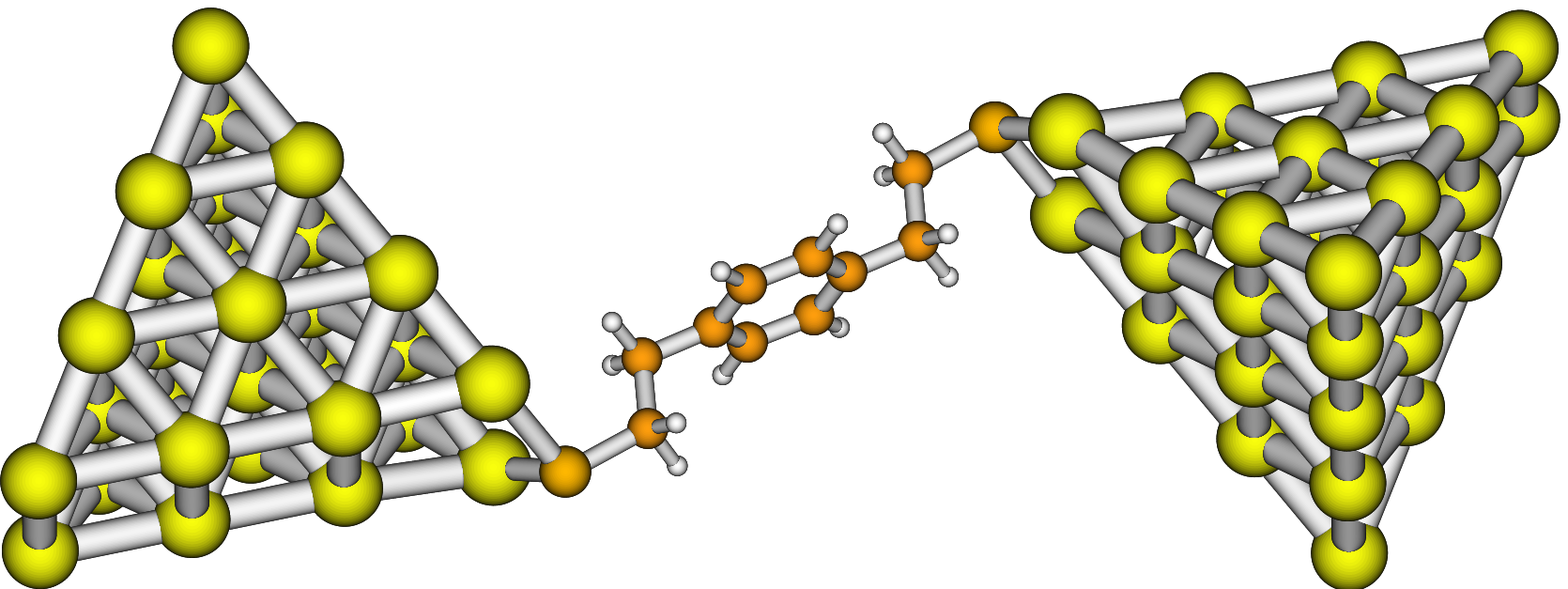}

\includegraphics[ width=8cm]{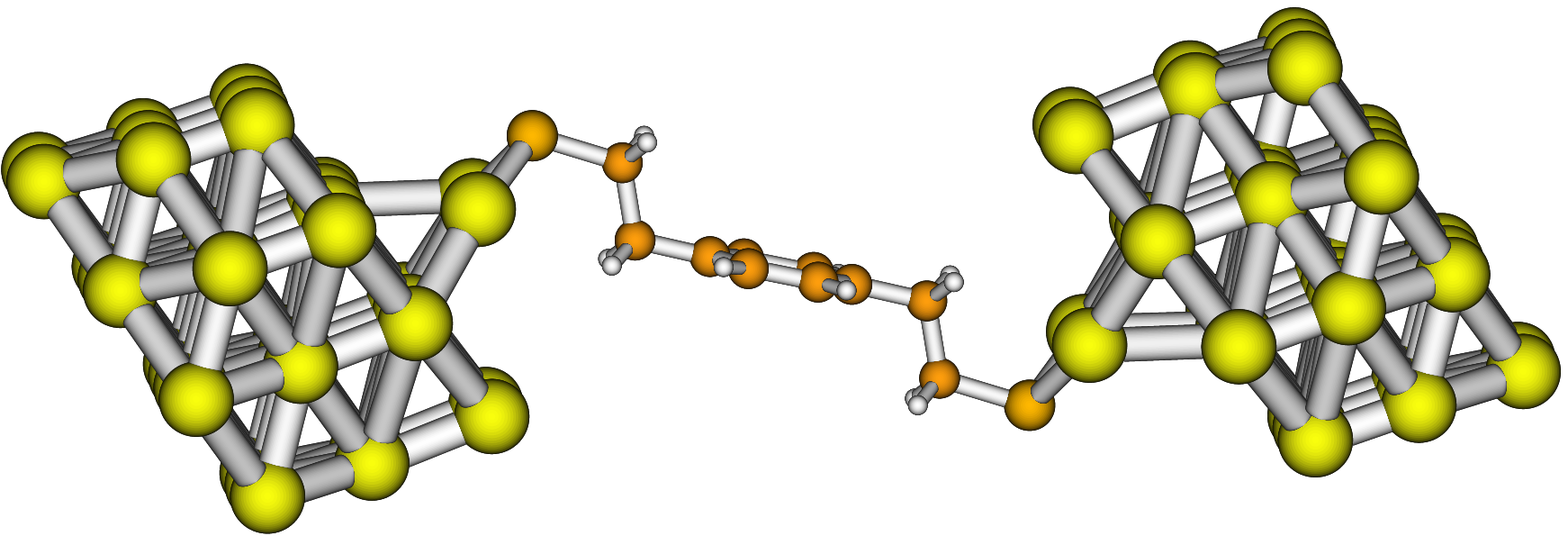}

\includegraphics[ width=8cm]{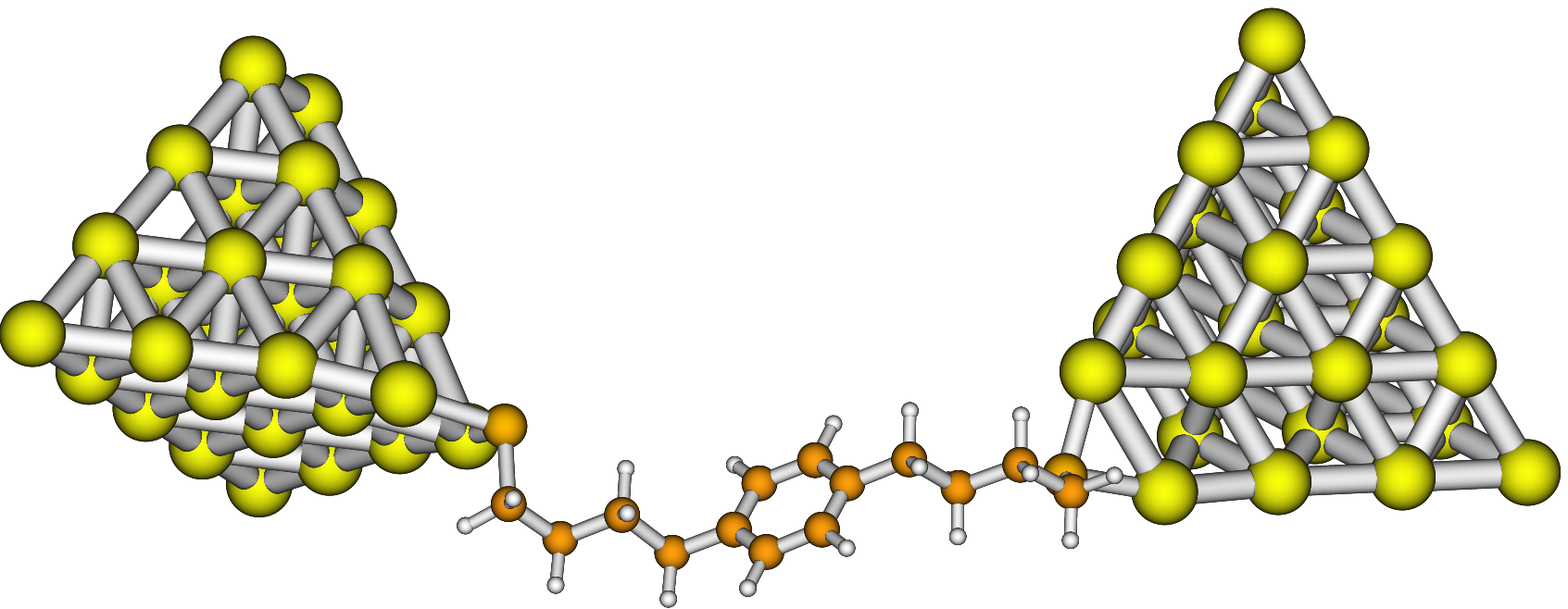}
\end{center}
\caption{The three molecular junctions investigated: 
benzene-di(ethanethiolate) (BDET) 
between pyramidal  gold contacts (top), BDET between cuboid-shaped gold contacts (middle), and
benzene-di(butanethiolate) (BDBT) between pyramidal  gold contacts (bottom).}
\label{modelsystem}
\end{figure}

The theoretical  methodology outlined above has been applied to study vibrational 
effects in electron transport through the three different molecular junctions depicted 
in Fig.\ \ref{modelsystem}, which include benzene-di(ethanethiolate) (BDET) 
between pyramidal  and cuboid-shaped gold  contacts as well as
benzene-di(butanethiolate) (BDBT) with a pyramidal contact geometry.
To reduce the computational effort, in the calculations presented below, 
the four vibrational modes with the strongest vibronic coupling
(as determined by the ratio of the electronic-vibrational coupling $\kappa$ and 
the electronic coupling $\Gamma$) 
were explicitly taken into account. 
Furthermore, the number of electronic states $| \phi_{m} \rangle$ on the molecular bridge, 
which were explicitly
included in the calculation, 
was reduced by including only those with energies in the vicinity of the Fermi energy 
(12 for BDET and 6 for BDBT).

\subsection{Benzenedi(ethanethiolate) with a pyramidal gold cluster geometry}\label{bdet_tip}

\begin{figure}
\begin{center}
\includegraphics[width=8cm]{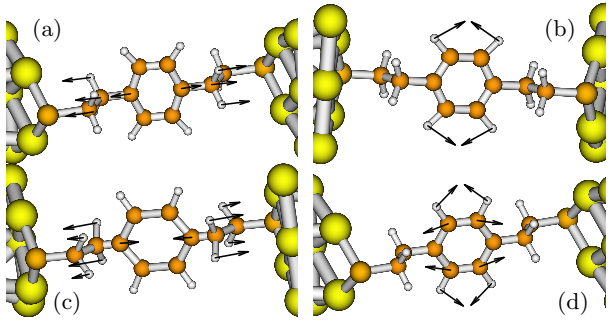}
\end{center}
\caption{Normal modes of BDET included in the calculation.}
\label{vibs_spacer}
\end{figure}

We first consider the transport characteristics of BDET between two pyramidal 
gold clusters (Fig.\,\ref{modelsystem}, top panel). 
Fig.\ \ref{vibs_spacer} depicts the four vibrational normal modes of BDET with the strongest 
vibronic coupling included in the calculation, which 
can be characterized as: C-C-C bending ((a) and (c)), C-C-H bending (b), and 
C-C stretching (d). The corresponding normal mode frequencies and vibronic 
coupling constants in the two 
most important molecular 
orbitals are given in Tab.\ \ref{kap_spacer}.
\begin{table}
\caption{Parameters of the four most important 
vibrational modes of BDET between pyramidal gold contacts 
including frequencies, periods, and vibronic coupling  in the electronic states A and B.
}
\begin{tabular}{c||c|c||c|c}
&$\omega$ (cm$^{-1}$)& T (fs) & $\kappa^{(A)}$ (meV)& $\kappa^{(B)}$ (meV)\\ 
\hline
 (a)&  544.48&	61&	 76 &  22	\\
 (b)& 1197.11&	28&	 51 &  69	\\
 (c)& 1229.49&	27&	110 &  47	\\
 (d)& 1671.56&	20&	136 & 162	
\end{tabular}
\label{kap_spacer}
\end{table}
\begin{figure}[h]
\begin{center}
\includegraphics[width=8cm]{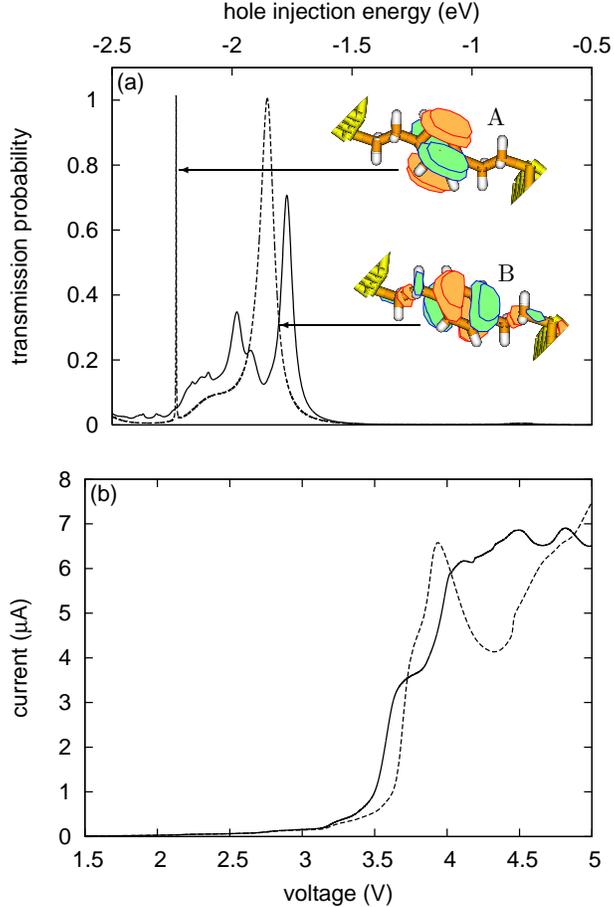}
\end{center}
\caption{(a)  Total transmission
probability through a BDET molecular junction with a pyramidal gold cluster geometry at zero 
voltage as a function of the initial energy of the hole (relative to the Fermi energy). 
The two orbitals, denoted A and B, dominate the transmittance at the indicated peaks. 
Only the energy range with non-zero transmittance in the interval 
[-2.5:+2.5], corresponding to a voltage window of
5 V, is shown.
(b) Current-voltage characteristic of BDET bound to pyramidal gold contacts. Shown are
results of calculations with (solid line) and without (dashed line) molecular vibrations.
Only the non-zero part of the positive voltage range is shown.
The different lines in the two panels show  results of calculations 
with (dashed line) and without (solid line) vibronic coupling.}
\label{t_spacer}
\end{figure}
%elastic transmission

The transmission probability of BDET at zero voltage is depicted in Fig.\,\ref{t_spacer} (a).
In addition to the transmission probability based on a vibronic calculation, also the
result of a purely electronic calculation (where all vibronic coupling constants $\kappa^{(m)}_l$
have been set to zero) is shown. The result of the purely electronic calculation exhibits
two pronounced resonance peaks  at energies -2.23\,eV 
and -1.85 eV. These two resonances are the closest to the Fermi energy
and thus determine the transport process. The two resonances are caused by the orbitals 
A and B depicted in Fig.\,\ref{t_spacer} (a).
Assuming that the projected molecular space includes the same number of electron pairs
as in the isolated molecule, the states A and B would correspond to the 
HOMO-4 and HOMO-5 of the isolated molecule.
While orbital A resembles an $e_{1g}$-orbital of benzene, orbital B has additional contributions 
on the ethyl-groups and the sulfur atoms. 
As a consequence, state A has smaller electronic coupling to the leads ($\Gamma_{AA} = 1.9\cdot
10^{-4}$ eV, corresponding to a lifetime of $\tau = \hbar/\Gamma_{AA} = 3466$ fs) 
resulting in a very narrow peak in 
the transmission probability, whereas the significant coupling of orbital B to the leads 
($\Gamma_{BB} = 1.1\cdot 10^{-1}$ eV, corresponding to a lifetime of $\tau=6$ fs) results in a 
rather broad structure. 
It is noted that in this estimate of the lifetime, 
the width function $\Gamma$ has been taken 
at the corresponding resonance peak energy. Due to the interaction with the leads and 
neighboring states,  the peak energy, in general, 
does not coincide with  eigenenergies of the states after projection, $E_m$.
For example, there exist four pairwise degenerate sulfur p-states with energies close to the
Fermi level, which do not appear themselves in the transmission probability but 
 influence the transmission  via their interaction with states A and B. 
This interaction is mediated  by the gold clusters via 
the non-diagonal elements in the self energy matrix.
%While the peak due to orbital A is not shifted with respect to the eigenenergy $E_A$, 
%the peak due to
%orbital B is shifted by -0.01\,eV. {\em this is very small, maybe should not be mentioned here}

The comparison between the results of vibronic (solid line in Fig.\,\ref{t_spacer} (a)) and purely 
electronic (dashed line) calculations demonstrates that the electronic-vibrational coupling in 
BDET alters the transmission probability noticeably. In particular, it results in a shift
of the electronic resonance peaks due to nuclear relaxation and the appearance of new structures,
which correspond to vibrational states in the molecular cation.
 Specifically, the electronic resonance of state B is shifted by  0.08 eV 
to the peak corresponding to the $0_0^0$ 
transition \cite{note2}  and the transmission shows 
resonance structures at -1.92 eV and -1.98 eV corresponding to  single vibrational excitations
 of mode (b) ($(\nu_b)_0^1$ transition) and (d) ($(\nu_c)_0^1$ transition).
It is noted that in contrast to previous  calculations \cite{Benesch06}, where 
charge transport through
BDET was described as electron instead of hole transport, 
the shift due to nuclear relaxation is in the 
opposite direction and the higher excited vibrational states in the  vibrational progressions
are located at lower energies.
The $0_0^0$ transition peak belonging to state A, which may be 
expected to be a narrow line, cannot be seen because it is coincidentally shifted into resonance 
with a sulfur p-state transmission peak. 
This peak, which does not appear directly in the transmission probability, 
causes a broadening of the resonance of state A. 
The structure resulting from this effect can be seen around peak A but it is too broad to assign
individual modes.

The current-voltage characteristic of BDET is shown in Fig.\,\ref{t_spacer} (b). 
Due to the symmetry of the junctions, the current fulfills the relation $I(-V) = -I(V)$ 
and, therefore, only
the positive voltage range is shown.
The result obtained from a purely electronic calculation (dashed line) shows an increase of the
current at about 3.5 V related to hole transport through state B. 
The increase is followed by a pronounced decrease of the 
current at 4 V. This negative-differential resistance (NDR) effect is a result of the voltage 
dependence of the self energies $\Sigma_L^{\rm lead}$, $\Sigma_R^{\rm lead}$ 
and the corresponding width  functions $\Gamma$ and will be discussed in more detail below.
The weakly coupled state A results only in a small step-like increase of the current shortly
before 4.5 V and the following rise of the current is due to the increasing
width of the resonance peak due to state B which appears at -1.85 eV 
in the transmission probability.

Including the coupling to the nuclear degrees of freedom changes the current-voltage 
characteristic significantly. 
In particular, the electronic-vibrational coupling results in an earlier onset of the current, 
a quenching of the NDR effect  as well as step-like substructures
which can be associated with vibrational excitation in the B state.
The earlier onset of the  current is due to the fact that upon 
interaction with vibrations the resonance peaks
are shifted to higher energies and thus enter the voltage window at lower voltage. If the
voltage window is large enough to comprise all vibronic features, the difference 
between purely electronic and vibronic results for the current 
vanishes.
The vibrational substructures are a result of the splitting of each 
electronic resonance into several vibronic resonances (cf. Fig.\,\ref{t_spacer} (a)), 
which contribute with different weights (determined by the respective Franck-Condon factors) 
to the transmission. 
In contrast to the purely electronic case, the current thus increases in several  
steps. This effect is well known from previous model studies \cite{Cizek04,Koch05}.

\begin{figure}
\begin{center}
\includegraphics[width=8cm]{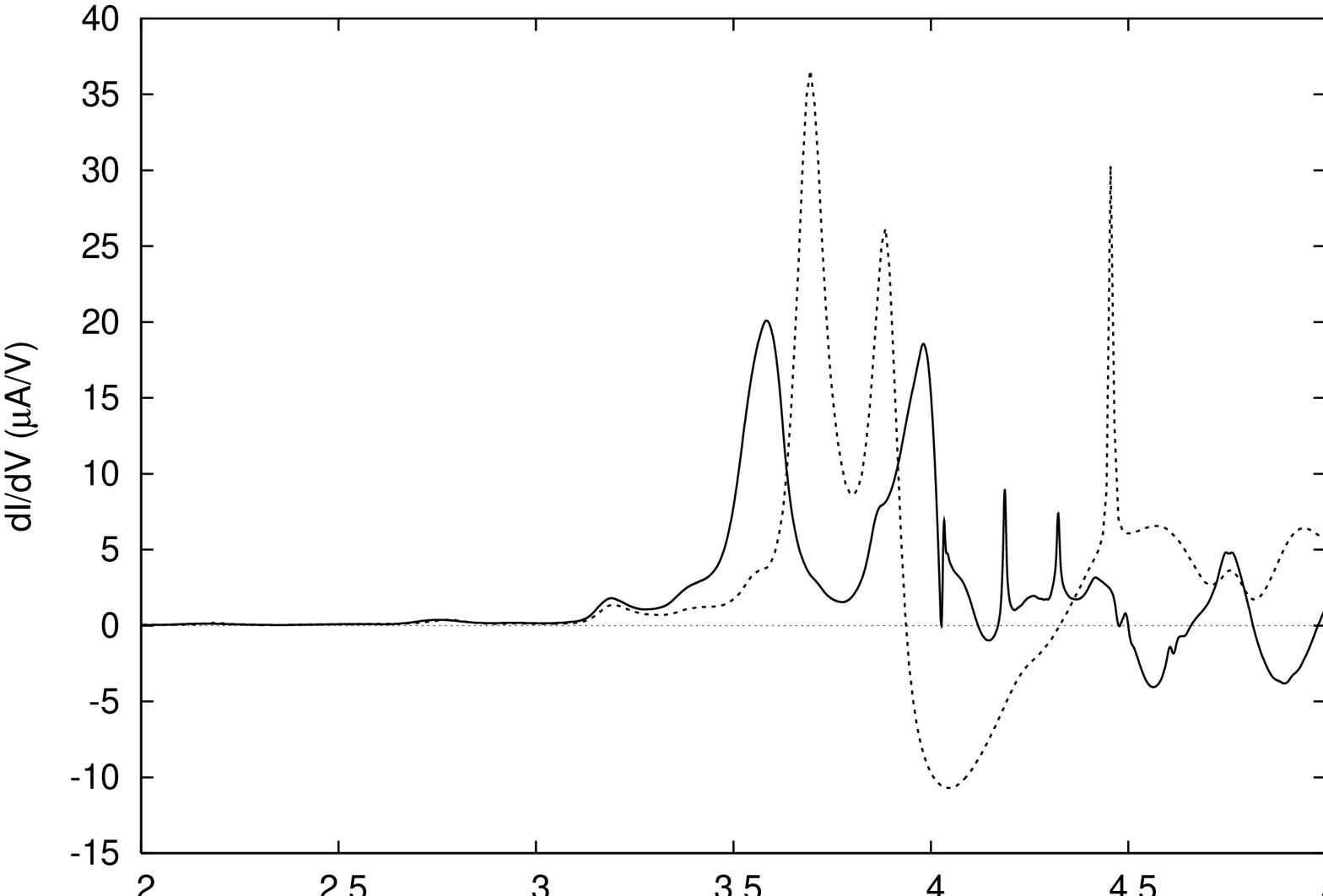}
\end{center}
\caption{Conductance of BDET based on a purely electronic (dashed line) and 
a vibronic (solid line) calculation.}
\label{di_dv}
\end{figure}

Smaller structures in the current-voltage characteristics are more clearly revealed in the
differential conductance, $dI/dV$.
Fig.\ \ref{di_dv} shows the differential conductance of BDET 
obtained by numerical differentiation of the
current in Fig.\ \ref{t_spacer} (b). Overall, the differential conductance reflects 
the transmittance
of the junction. However, in contrast to the
transmission probability at zero voltage (depicted in Fig.\ \ref{t_spacer} (a)), it also incorporates 
its voltage dependence and the effects of the Fermi distribution in the leads. In the purely 
electronic conductance (dashed line), this difference 
manifests itself in the splitting of the peak due to 
state B as well as the negative differential resistance between 4 and 4.3 Volt. At 4.5 V a 
narrow resonance due to state A can be seen.
The differential conductance based on the vibronic  calculation exhibits
pronounced structures  due to vibronic coupling. These include the shifted onset 
(corresponding to the $0_0^0$ transitions) of state A at 3.58 V and state B at 4.27 V
as well as resonance  peaks at 3.98 V corresponding the $(\nu_d)_0^1$ vibrational excitation in state B 
and  at 4.32 V reflecting the $(\nu_a)_0^1$
excitation in state A. 
The differential conductance based on the vibronic calculation  shows, furthermore,
small  regions of negative differential resistance for larger voltages,  which do
not exist in the purely electronic differential conductance.

\begin{figure}
\begin{center}
\includegraphics[width=8cm]{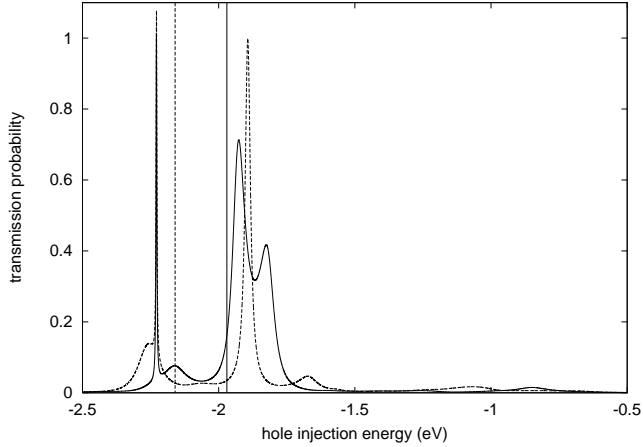}
\end{center}
\caption{Transmission of BDET at voltages 3.94 V (solid line) and 4.32 V (dashed
line) based on a purely electronic calculation. 
The two vertical lines indicate the respective lower integration boundaries.}
\label{t_at_v}
\end{figure}

Finally, we consider the mechanisms responsible for the NDR effect observed in the results 
discussed above.
As was mentioned above, the NDR effect is caused by the  voltage 
dependence of the self energies. These result in a voltage dependence of
the transmission probability which in turn determine
the current via the integral in Eq.\ (\ref{current}).
To elucidate the mechanism causing NDR in the purely 
electronic calculation (dashed lines in Figs.\ \ref{t_spacer} (b), \ref{di_dv}),
Fig.\ \ref{t_at_v} depicts   
the transmission probability for  voltages of 3.94 V (corresponding to the local 
maximum in the I-V curve) and 4.32 V (the local minimum in the I-V 
curve). The peak caused by orbital B in the  transmission for 3.94 V (dashed line) 
is much thinner than 
the corresponding peak in the  curve for 4.32 V (solid line). 
Although the current in the former case corresponds to the integral 
with a lower boundary (see vertical line in Fig.\ \ref{t_at_v}), 
the area under the '4.32 V'-transmission-curve is much smaller thus
explaining the NDR in the current-voltage characteristic and the conductance.
It should be emphasized that this NDR effect can only be described, if the energy and voltage 
dependence of the self energies is taken into account and will be missed within the often used 
wide-band approximation.

%\begin{figure}
%\begin{center}
%\includegraphics[width=8cm]{figures/t_4_5.eps}
%\end{center}
%\caption{Purely electronic transmission of BDET employing a 4-layer (dashed line) and 5-layer 
%(solid line) gold contact.}
%\label{t_4_5}
%\end{figure}
%%
%To justify the usage of this four layer cluster, we added a fifth layer and compared the 
%zero-voltage transmission function of those two systems.
%They agreed very well after taking into account the different Fermi levels and different shifts
%of the localized molecular states after projection (see Fig.\,\ref{t_4_5}). However, it should be 
%mentioned that the differences become more pronounced, if the transmission at non-zero voltages 
%is regarded.
%% 4 layer: ef=-5.147866\,eV; 5 layer: ef=-4.657326\,eV
%% difference for state D: -.117717\,eV; difference for state E: -.135512\,eV

\subsection{Benzenediethanthiolate with a cuboid-shaped gold-cluster geometry}\label{bdet_cub}

To investigate the influence of the gold-cluster geometry on the transport properties of the 
molecular junction, we have studied a system where the molecule BDET is bound to 
gold-clusters with different geometry of cuboid shape 
(cf. Fig.\,\ref{modelsystem} middle panel). 
To build this geometry we started from the optimized two-layer system described above.
Next, the number of gold atoms in the second layer was increased from 5 to 12  and two more 
gold-(111)  layers consisting of 12 gold atoms each were added, thus 
resulting in a metal cluster of 38 gold atoms 
on each side. For better comparison with the  results obtained in Sec. \ref{bdet_tip}, the 
same value was used for the Fermi energy. The  four normal modes 
and the molecular electronic states included
in the calculation are very similar to those in Sec.~\ref{bdet_tip}.

\begin{figure}
\begin{center}
\includegraphics[width=8cm]{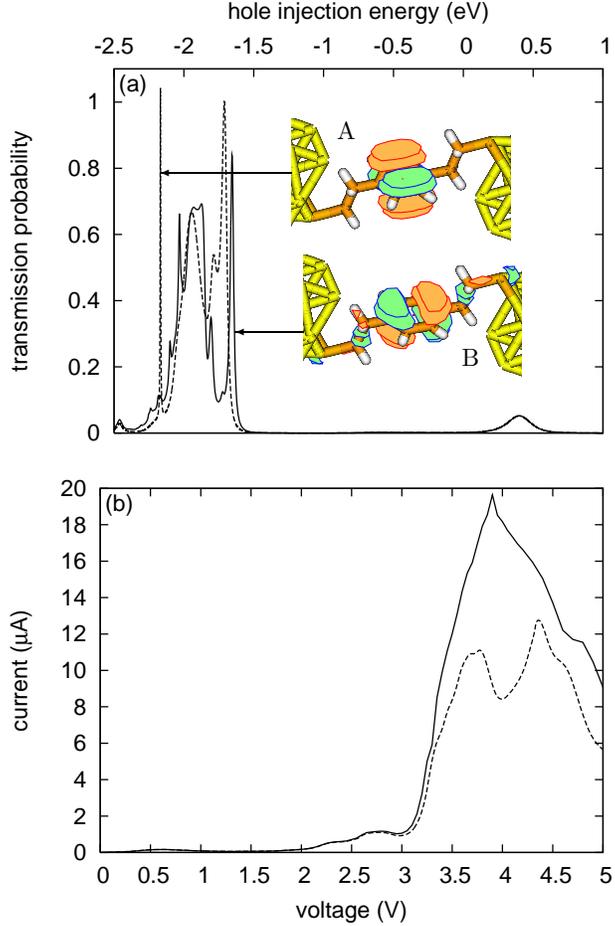}
\end{center}
\caption{(a)  Total transmission
probability through a BDET molecular junction with a cuboid-shaped gold cluster geometry at zero 
voltage as a function of the initial energy of the hole (relative to the Fermi energy). 
The two orbitals, denoted A and B, dominate the transmittance at the indicated peaks. 
Only the energy range with non-zero transmittance in the interval 
[-2.5:+2.5], corresponding to a voltage window of
5 V, is shown.
(b) Current-voltage characteristic of BDET bound to cuboid-shaped gold contacts. Shown are
results of calculations with (solid line) and without (dashed line) molecular vibrations.
Only the non-zero part of the positive voltage range is shown.
The different lines in the two panels show  results of calculations 
with (dashed line) and without (solid line) vibronic coupling.}
\label{t_plains}
\end{figure}

Fig.\ \ref{t_plains} (a) shows the zero-voltage transmission probability for this system.
The comparison with the corresponding result for the pyramidal gold-cluster geometry (Fig.\ 
\ref{t_spacer}) reveals overall a significant effect of the cluster geometry. Considering 
specifically the
transmission based on the purely electronic calculation, 
the resonance due to state B is split into three  peaks at 
-1.71 and -1.79 and -1.95 eV. This is caused by the electrode-mediated 
interaction of state B with both lower and higher
lying states, which is more pronounced in the cuboid gold-cluster geometry.
In addition, there is a small broad peak at 0.39 eV due to two sulfur p-orbitals,
which is missing in the pyramidal geometry. The resonance due to state A in the 
zero-voltage transmission, on the other hand,  is almost unaffected by the 
geometry of the gold cluster. This is to be expected
because state A is exclusively localized on the benzene ring.  
The coupling of states A and B to the contacts is of 
comparable size as for the pyramidal gold clusters: 
$\Gamma_{AA}(-2.17) =  4.0\cdot 10^{-4}$\,eV (corresponding to a lifetime of $\tau=1646$ fs),
$\Gamma_{BB}(-1.98) =  6.5\cdot 10^{-2}$\,eV ($\tau=10$ fs), and
$\Gamma_{BB}(-1.71) =  1.1\cdot 10^{-1}$\,eV ($\tau=6$ fs).

Considering the results of the vibronic calculations (full line in 
Figs.\ \ref{t_spacer}, \ref{t_plains}),
the most significant difference between the two cluster geometries
is the appearance of the $0_0^0$-transition 
and vibrational progressions due to state A. These were not visible
in the system with the pyramidal gold clusters due to accidental coincidences with transmission
features of other states. In the system with cuboid cluster geometry, the $0_0^0$-line
due to state A appears at -2.03\,eV and the $(\nu)_0^1$
transition peaks corresponding to all four vibrational modes  can be seen at -2.10 
(a), -2.17 (b), -2.18 (c), and -2.24 (d) eV.

Fig.\ \ref{t_plains} (b) shows the current-voltage characteristic for the BDET system 
with cuboid gold clusters.
The current-voltage characteristic based on the purely electronic calculation (dashed line) has a 
similar appearance as in the system with the pyramidal cluster (cf.\ Fig.\ \ref{t_spacer} (b)). 
Noticeable differences are the earlier onset and the overall larger value of the current as well as 
the appearance of two (instead of one)  maxima with corresponding regions of NDR.
The larger current is a manifestation of the overall larger molecule-lead coupling provided 
by the binding to the cuboid shaped gold cluster. The earlier onset of the current 
is caused by orbitals localized on the
sulfur atoms. The appearance of two maxima is due to the splitting of the resonance
peak of state B. 
We next consider the current based on the vibronic calculation 
(full line in Fig.\ \ref{t_plains} (b)).  In contrast to the system with the pyramidal gold custer 
geometry, 
vibrational substructures are less pronounced and the current obtained with the vibronic 
calculation is in some regions significantly larger than that based on the electronic current.
Furthermore, NDR effects are only partly quenched by the interaction with the 
vibrational degrees of freedom.

%The question arises how the vibronic current can be larger than that without vibrations?
%The vibronic coupling increases the transmission probability by shifting and splitting the
%resonance peaks. Thus, the resonances may reach positions where the electronic coupling to the
%leads, but also to other molecular states via the leads, is significantly different from the
%electronic interaction at the initial position. Consequently, the transmission probability may 
%change and in this case increases.

To conclude this section, the comparison of the data for BDET bound to two different gold clusters 
shows that the geometry of the gold contact may have a 
significant influence on the transport properties.
This is in accordance with results of Ref.\ \onlinecite{Ke2005}, which show that
the  transport characteristics of benzenedithiol depends on the surface structure and the lateral 
size  of the electrodes.
Our results show that the influence of the gold contact geometry is particularly pronounced
 for transport through molecular orbitals which have a significant 
contribution at the boundary to the contacts such as state B in the system studied here. 
In experiments, where  it is often not straightforward to control  the molecule-lead binding geometry,
this difference may result in fluctuations of the current for different realizations of
molecular junctions.

\subsection{Benzenedi(butanethiolate)}

\begin{figure}
\begin{center}
\includegraphics[width=12cm]{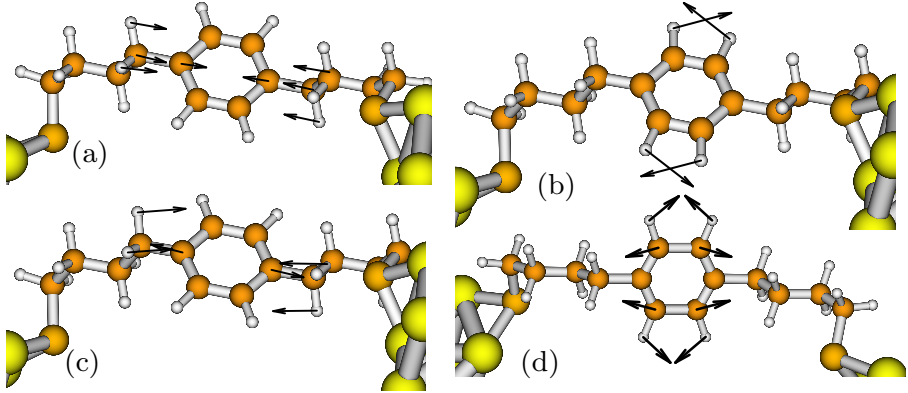}
\end{center}
\caption{Normal modes of BDBT included in the calculation.}
\label{vibs_bdbt}
\end{figure}

\begin{figure}
\begin{center}
\includegraphics[width=8cm]{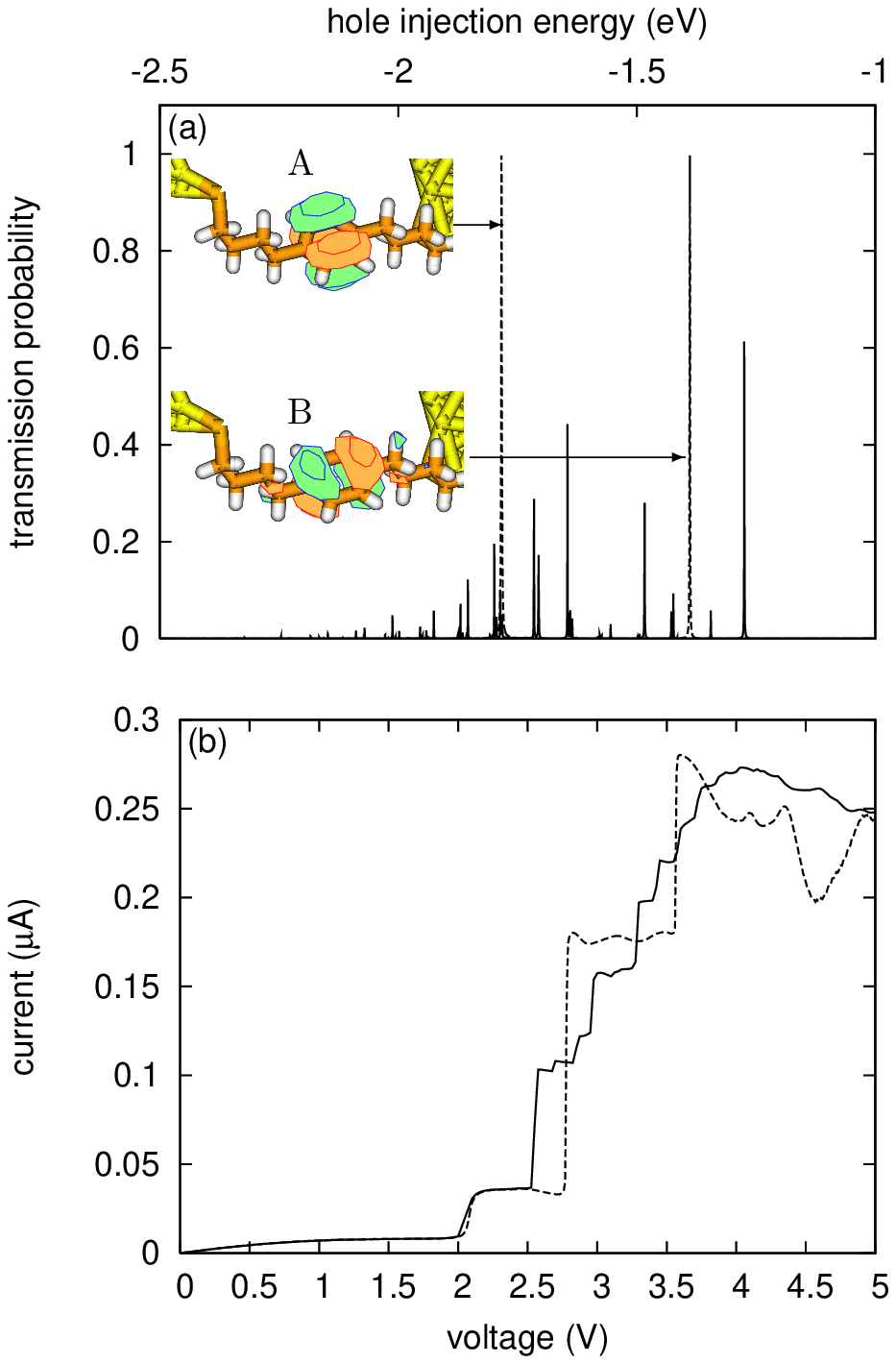}
\end{center}
\caption{(a) Total transmission
probability through a BDBT molecular junction with a pyramidal gold cluster geometry at zero 
voltage as a function of the initial energy of the hole (relative to the Fermi energy). 
The two orbitals, denoted A and B, dominate the transmittance at the indicated peaks. 
Only the energy range with non-zero transmittance in the interval 
[-2.5:+2.5], corresponding to a voltage window of
5 V, is shown.
(b) Current-voltage characteristic of BDBT bound to pyramidal gold contacts. Shown are
results of calculations with (solid line) and without (dashed line) molecular vibrations.
Only the non-zero part of the positive voltage range is shown.
The different lines in the two panels show  results of calculations 
with (dashed line) and without (solid line) vibronic coupling.}
\label{t_long}
\end{figure}

The systems studied in the previous sections are characterized by a moderate molecule-lead coupling,
corresponding to a relatively short lifetime of the electron on the molecular bridge. 
Accordingly, vibrational effects are noticeable but not very pronounced.
The coupling between the $\pi$-system of the molecular bridge and the gold electrodes
can be reduced (and thus the lifetime of the electron on the bridge extended) 
by increasing the length of the alkyl-spacer group. As an example of such a system, we consider
in this section electron transport through p-benzene-di(butanethiolate) (BDBT) 
(cf.\ Fig.\ \ref{modelsystem}, bottom panel). 
Thereby, the contacts were modelled   with a  gold cluster  of pyramidal geometry. 
The transport properties
of this system  can be well described including six 
electronic states localized at the molecular bridge in the calculation. 
These comprise the states A and B illustrated in Fig.\ \ref{t_long}, which are similar as in BDET, 
and four sulfur p-orbitals. As in the previous system, from all 54 totally symmetric normal modes
the four with the strongest vibronic coupling were included explicitly in the calculation.
The modes are depicted in Fig.\ \ref{vibs_bdbt}.
Their frequencies and coupling  constants are given in Table \ref{kap_bdbt}.

\begin{table}
\caption{Parameters of the four most important 
vibrational modes of BDBT between pyramidal gold contacts 
including frequencies, periods, and vibronic coupling  in the electronic states A and B.}
\begin{tabular}{c||c|c||c|c}
&$\omega$ (cm$^{-1}$)& T (fs)& $\kappa^{(A)}$ (meV)& $\kappa^{(B)}$ (meV)\\ 
\hline
 (a)&  566.12 &	59&  69 &  29	\\
 (b)& 1198.40 &	28&	 52 &  73	\\
 (c)& 1232.90 & 27&	114 &  56	\\
 (d)& 1676.10 &	20&	132 & 170	
\end{tabular}
\label{kap_bdbt}
\end{table}

Fig.\ \ref{t_long} (a) shows the zero-voltage transmission probability of BDBT. In contrast to BDET,
the transmission probability exhibits well-separated narrow resonance structures. 
This is due to the significantly smaller electronic coupling.
The results based on the purely electronic calculation exhibit two 
narrow resonances at -1.78 and
-1.39 eV. These are due to states A and B, respectively, which couple to the gold electrodes with
coupling strengths of $\Gamma_{AA}(-1.78)=1.0\cdot 10^{-3}$\,eV ($\tau=659$ fs) and 
$\Gamma_{BB}(-1.39)=5.4\cdot10^{-4}$\,eV ($\tau=1220$ fs), respectively. Due to the small
coupling, these states exhibit negligible level shift. 
Including vibronic coupling both peaks are shifted to the respective $0_0^0$ transition
and pronounced vibrational progressions appear. The individual peak heights correspond 
to the respective Franck-Condon factors. Table \ref{trans_spec} gives an assignment
of the most prominent resonances.

\begin{table}[h]
\begin{center}
\begin{tabular}{c||c|c|}
orbital A& ~~$(\nu)_0^1$~~& ~~$(\nu)_0^2$~~\\ \hline
mode (a)& -1.72& -1.79\\
mode (b)&	&	\\
mode (c)& -1.80& -1.93\\
mode (d)&	-1.85& -2.07\\ \hline\hline
orbital B &$(\nu)_0^1$&$(\nu)_0^2$\\ \hline
mode (a)& -1.35& 	\\ 
mode (b)& -1.42& -1.56  \\
mode (c)& -1.43& 	\\
mode (d)& -1.48& -1.71\\
\end{tabular}\end{center}
\caption{Assignments of the most pronounced  vibronic transitions in the transmission function  of BDBT
depicted in Fig.\ \ref{t_long} (a). The location of the features are given in eV.  Blank entries correspond to 
transitions that could not be identified because of too small
vibronic coupling.}
\label{trans_spec}
\end{table}

The current-voltage characteristic of BDBT is depicted in \ref{t_long} (b).
The result based on the purely electronic calculation shows 
two prominent steps corresponding to states A and B at voltages $2.76$ V and 
$3.56$ V, respectively. 
In addition, there is a small current for low voltages and another step-like structure
at 2 V. Both features are due
to sulfur p-states, which cause a small but very broad peak in the transmission probability
around the Fermi level and a small, narrow peak at -1\ eV, 
which can be seen in the transmission probability at 2 V only (data not sown), but
not in the zero-voltage transmission.
As in the other model systems, NDR effects caused by the same mechanism as for BDET 
appear at larger voltages.
Upon including vibronic coupling, the onset of the current related to states A and B is 
shifted to lower voltage due to nuclear relaxation. 
Furthermore, a number of additional steplike structures appear which
can be assigned to excitation of the different vibrational modes in states A and B.
The most pronounced features are analyzed in Table \ref{cur_spec}.
%The next step, due to orbital B without
%vibrational excitation, occurs at 2.55\,V and is followed by a step at 2.68\,V due to single
%excitation of mode (a), a step at 2.85\,V due to single excitations of modes (b) and (c), and a
%somewhat larger step at 2.95\,V due to the excitation of mode (d). At 3.27\,V transmission through
%orbital A starts to contribute to the current and is followed by a step at 3.42\,V due to the 
%$(a)_0^1$ transition in orbital A and the $(d)_0^2$ transition of mode (d) in orbital B. There 
%are two more steps that
%can be uniquely assigned to transitions. The step at 3.57\,V, which is caused by single
%vibrational excitations of modes (b) and (c) in orbital A, and the step at 3.73\,V, which is due to
%excitation of mode (d) in orbital A. 
The NDR following the last
step still exists but is significantly smaller compared to the purely electronic case.

\begin{table}[h]
\begin{center}
\begin{tabular}{c||c|c|}
 & orbital A& orbital B\\ \hline
 0$_0^0$ & 3.27 & 2.55 \\
 $(\nu_a)_0^1$ & 3.42	& 2.68	\\
 $(\nu_b)_0^1$ & 3.57	& 2.85	\\
 $(\nu_c)_0^1$ & 3.57	& 2.85	\\
 $(\nu_d)_0^1$ & 3.73	& 2.95	\\
 $(\nu_d)_0^2$ &	& 3.42	\\
\end{tabular}\end{center}
\caption{Assignments of the most pronounced  vibronic transitions in the I-V curve 
of BDBT depicted in Fig.\ \ref{t_long} (b). The location of the features are given in V. 
 Blank entries correspond to transitions that could not be identified, because of too small
vibronic coupling.}
\label{cur_spec}
\end{table}

%% file: conclusions.tex
\section{Conclusions}

In this paper we have studied the effect of vibrational motion on
resonant charge transport in single molecule junctions. 
The methodology used to describe vibrationally coupled charge transport is based on a 
combination of first-principles electronic structure calculations to
characterize the system and inelastic scattering theory. 
To apply this methodology to systems where the transport is dominated by the occupied molecular orbitals
of the junction, we have extended the 
approach to allow for hole transport.

We have applied the methodology to molecular junctions 
where a benzene molecule is connected via alkanethiol
bridges to two gold electrodes.
This class of molecular junctions was chosen, because
the change of the length of the alkyl chain 
allows a systematic variation of the coupling between the $\pi$ system of the phenyl ring and 
the gold electrodes and thus a variation of
the lifetime of the electron on the molecular bridge.
The ratio between the vibronic coupling 
and the electronic molecule-lead coupling determines the
importance of vibronic effects in molecular junctions.
The results of this study demonstrate this trend. 
Our previous work \cite{Benesch06} had indicated that
in benezendithiolate, which is characterized by rather strong molecule-lead
coupling (and thus a very short lifetime of the electron on the molecular bridge), 
vibrational effects in resonant transport are almost negligible.
Benezendibutanthiolate, the system with the smallest molecule-lead
coupling investigated, exhibits pronounced vibronic effects. 
In particular, electronic-vibrational coupling result in a splitting of electronic resonances into
vibronic subresonances in the transmission probability.
In the current-voltage characteristic, 
vibronic effects manifest themselves in steplike structures, which can
be associated with vibrational states in the molecular cation.
Benezendiethanethiolate  is an intermediate case, which shows noticeable 
though not very pronounced vibronic effects.
The study also shows that electronic-vibrational coupling may result in 
a quenching of negative differential resistance effects, which are caused by the voltage dependence
of the self energies,
and a significantly 
altered overall  magnitude of the  current.

%It turned out, that vibronic coupling may lead to quenching of NDR-features or enhancement of the
%current. The NDR itself is due to the voltage-dependence of the transmission function, while the
%enhancement occurs because, on vibronic interaction, electronic resonances are shifted towards 
%positions where the electronic coupling is larger.

%% file: paper.bbl
\begin{thebibliography}{10}
\bibitem{note2}
Here, we are using spectroscopic notation $(\nu)_n^l$ to denote vibronic
  transitions of mode $\nu$ from the $n$th vibrational state in the neutral
  molecule to the $l$th vibrational state in the molecular cation. $0_0^0$
  denotes the transition between the respective vibrational ground states.
\end{thebibliography}
